\documentclass{amsart}

\usepackage{amsmath}
\usepackage{csquotes}
\usepackage{amssymb}
\usepackage{amsthm}
\usepackage{geometry}
\usepackage[numbers,sort&compress]{natbib}
\usepackage{hyperref}
\usepackage{booktabs}
\usepackage{caption}
\usepackage{siunitx}
\usepackage{tabularx}
\usepackage{graphicx}
\usepackage{adjustbox}
\usepackage{multirow}
\usepackage{csquotes}
\usepackage{amsmath}
\newtheorem{theorem}{Theorem}[section]
\newtheorem{lemma}[theorem]{Lemma}
\theoremstyle{definition}

\newtheorem{example}[theorem]{Example}
\theoremstyle{conjecture}
\newtheorem{conjecture}[theorem]{Conjecture}

\theoremstyle{proposition}
\newtheorem{proposition}[theorem]{Proposition}
\theoremstyle{remark}
\newtheorem{remark}[theorem]{Remark}

\theoremstyle{corollary}

\numberwithin{equation}{section}

\begin{document}

	\title{Linear codes using simplicial complexes}

	
	
	
	
	\author{Vidya Sagar}
	\address{(V. Sagar)Department of Mathematics, IIT Delhi, New Delhi, India 110016}
	
	\email{vsagariitd@gmail.com}
	
	
	\author{Ritumoni Sarma}
	\address{(R. Sarma)Department of Mathematics, IIT Delhi, New Delhi, India 110016}
	
	\email{ritumoni407@gmail.com}
	
	
	
	
	
	\begin{abstract}
	Certain simplicial complexes are used to construct a subset $D$ of $\mathbb{F}_{2^n}^m$ and $D$, in turn, defines the linear code $C_{D}$ over $\mathbb{F}_{2^n}$ that consists of $(v\cdot d)_{d\in D}$ for $v\in \mathbb{F}_{2^n}^m$. Here we deal with the case $n=3$, that is, when $C_{D}$ is an octanary code. We establish a relation between $C_{D}$ and its binary subfield code $C_{D}^{(2)}$ with the help of a generator matrix. For a given length and dimension, a code is called distance optimal if it has the highest possible distance.	With respect to the Griesmer bound, five infinite families of distance optimal codes are obtained, and sufficient conditions for certain linear codes to be minimal are established.

	\end{abstract}

	\maketitle

	
	\section{INTRODUCTION}
	
	Throughout the manuscript, $\mathbb{F}_{q}$ is the field with $q$ elements; however, in this, we work over $\mathbb{F}_{2^3}$ and more generally $\mathbb{F}_{2^n}$ for $n\in \mathbb{N}$. Given a code $C$ over $\mathbb{F}_{q^n}$, the code $C^{(q)}$ over $\mathbb{F}_{q}$ is constructed using the trace map in \citep{Ding}. The code $C^{(q)}$ is referred to as subfield code with respect to $C$. Hyun et al. \citep{Hyun_Lee} studied linear codes by choosing some specific defining set obtained using simplicial complexes and produced certain families of binary optimal linear codes. In \citep{Hyun_Kim}, Hyun et al. used posets in place of simplicial complexes to study linear codes. They constructed binary linear codes that are minimal as well as optimal, but the Ashikhmin-Barg condition \citep{suffConMinimalcode} does not hold for them. In 2021, Zhu et al. studied codes over $\mathbb{F}_{4}$ and produced minimal optimal codes. In \citep{Wu_Li}, the authors studied linear codes over $\mathbb{F}_{4}$ and the corresponding subfield codes, and obtained their weight distributions. They used simplicial complexes to construct defining set and produced two infinite families of optimal linear codes. The weight distribution of a linear code is an important research topic in coding theory because it can be used to calculate the error correcting capability as well as the error probability of error detection and correction \citep{Ding_Ding, Dinh_Li, Klove}. The minimum Hamming distance of linear codes are well known for their importance in determining error-correcting capacity. As a result, finding optimal linear codes has become one of the central topics for researchers. A lot of work has been done in the recent past. In \citep{Hyun_Lee}, Hyun et al. shows how optimal codes can be utilized for the construction of secret sharing schemes having nice access structures following the framework discussed in \citep{YuanDing}. The secret sharing schemes are essential nowadays; in fact, these are used heavily in electronic voting systems, cryptographic protocols, banking systems etc. Minimal codes are helpful to construct the access structure of the secret sharing schemes \citep{Ding_Ding, MasseyMinimal}.\par 
	The study of minimal codes is an active research topic nowadays because of its applications in secure two-party computation and secret sharing schemes \citep{CarletDing, Chabanne_two_party, YuanDing, ShamirSSS}. This special class of codes is also important as these can be decoded by using the minimum distance decoding rule \citep{suffConMinimalcode}. Normally, it is difficult to identify all the minimal codewords of a given code even over a finite field of characteristic $2$. So researchers started looking into minimal codes.\par 
	Motivated by the above research, we in this manuscript, study octanary linear codes whose defining set is obtained from certain simplicial complexes. We also investigate binary subfield codes with respect to these codes and obtain their weight distributions. If defining sets are properly chosen, these codes are minimal and distance optimal. Up to some extent, we generalize these results to codes over $\mathbb{F}_{2^n}$. We also propose a few conjectures that generalizes results on octanary codes. We give a few examples that support our results.\par
	We arrange the remaining part of the manuscript as follows. Preliminaries are presented in the next section. With the help of simplicial complexes, octanary linear codes are investigated in section \ref{section_3}. We obtain sufficient conditions for some of these codes to be minimal. In this section, we prove these codes are distance optimal if the defining sets are chosen properly. Section \ref{section_4} studies subfield codes. We prove these codes are Griesmer code and obtain sufficient condition for these codes to be minimal. Weight distributions of these codes are also obtained in this section. Section \ref{section_5} concludes this manuscript.	
		
    \section{Definitions and Preliminaries}
    Throughout the manuscript, for $m\in \mathbb{N}$, we shall write $[m]$ to denote the set $\{1, 2,\dots ,m\}$.
    Suppose $C$ is an $[n,k,d]$-linear code over $\mathbb{F}_{q}$, where $n,k,d\in \mathbb{N}$. Denote by $A_i$, the cardinality of the set that contains all codewords of $C$ having (Hamming) weight $i$, $0\leq i\leq n$. Then the string $(1, A_1,\dots ,A_n)$ is called the \textit{weight distribution} of $C$. In addition, if the total number of $i\geq 1$ such that $A_i\neq 0$ is $l$, then $C$ is called an $l$-\textit{weight linear code}. The code $C$ is called \textit{distance optimal} if there exist no $[n,k,d+1]$-linear code (see\citep{wchuffman}). Next we recall the Griesmer bound.   
   \begin{lemma}\label{griesmerbound}\citep{Griesmer}
   	\textnormal{(Griesmer Bound)} If $C$ is an $[n,k,d]$-linear code over $\mathbb{F}_q$, then we have
   	\begin{equation}\label{GriesmerBound}
   		\sum\limits_{i=0}^{k-1}\left\lceil \frac{d}{q^i}\right\rceil \leq n,
   	\end{equation}  	
   where $\lceil \cdot \rceil$ denotes the ceiling function.
   \end{lemma}
    A linear code is called a \textit{Griesmer code} if equality holds in Equation \ref{GriesmerBound}. Note that every Griesmer code is distance optimal.
    
    \subsection{Construction of linear codes}
    For any $n, m\in \mathbb{N}$, let $D=\{d_1<d_2<\dots <d_n\}$ be a subset of $\mathbb{F}_{q}^m$ of cardinality $n$. Define \citep{wchuffman},
    \begin{equation}\label{genericConstruction}
    		C_{D}=\{(v\cdot d_1, v\cdot d_2, \dots , v\cdot d_n): v\in \mathbb{F}_{q}^m\}    	
    \end{equation}
    where $x\cdot y=\sum\limits_{i=1}^{m}x_iy_i$ for $x,y\in \mathbb{F}_q^m$.\\ The ordered set $D$ is termed as the \textit{defining set} of $C_{D}$. Note that on changing the ordering of $D$ we will get a permutation equivalent code. If $D$ is chosen appropriately, $C_{D}$ may possess good parameters. For $w\in \mathbb{F}_{q}^m$, the set
    \begin{equation*}
    	\textnormal{Supp}(w)=\{ i\in [m]: w_i\neq 0\}
    \end{equation*}
    is called the \textit{support} of $w$. Note that the Hamming weight of $w\in \mathbb{F}_q^m$ is $wt(w)=|\textnormal{Supp}(w)|$.\par
    For $v,w\in \mathbb{F}_{q}^m$, one says that $v$ covers $w$ if $\textnormal{Supp}(w)\subseteq \textnormal{Supp}(v)$. If $v$ covers $w$ we write $w\preceq v$. Let $C$ be a linear code over $\mathbb{F}_{q}$. An element $v_{0}\in C\setminus \{0\}$ is called \textit{minimal} if $v\preceq v_{0}$ and $v\in C\setminus \{0\}$ imply $v=\lambda v_{0}$ for some $\lambda\in \mathbb{F}_q^{\ast}$. If every nonzero codeword of a code is a minimal codeword then the code is termed as a \textit{minimal code}. Now we recall a result from \citep{suffConMinimalcode} regarding a sufficient condition for a linear code to be minimal.    
    \begin{lemma}\label{minimal_lemma}
    	\citep{suffConMinimalcode}\textnormal{(Ashikhmin-Barg)}
    	If the ratio of the minimum and the maximum (Hamming) weights of nonzero codewords in a linear code $C$ over $\mathbb{F}_{q}$ is larger than $\frac{q-1}{q}$, then $C$ is a minimal code.
    \end{lemma}
    Consider the map $\phi: \mathbb{F}_2^m\longrightarrow 2^{[m]}$ is defined as $\phi(w)=\textnormal{Supp}(w)$, where $2^{[m]}$ denotes the power set of $[m]$. Note that $\phi$ is a bijective map.\\
    A subset $\Delta$ of $\mathbb{F}_{2}^m$ is called a \textit{simplicial complex} if $u\in \Delta$ then $w\in \mathbb{F}_{2}^m \implies w\in \Delta$ whenever $w\preceq u$. An element $v\in \Delta$ is called \textit{maximal} if $v\preceq w$ and $w\in \Delta$ imply $v=w$. Let $L\subseteq [m]$. The simplicial complex generated by $L$ is denoted by $\Delta_{L}$ and is defined as 
    \begin{equation}
    	\Delta_{L}=\{w\in \mathbb{F}_{2}^m| \textnormal{ Supp}(w)\subseteq L\}.
    \end{equation}
    Note that $|\Delta_{L}|=|2^L|=2^{|L|}$.\par
    Given a subset $P$ of $\mathbb{F}_{2}^m$, define the polynomial (referred to as  m-variable generating function, see \citep{Chang}) $\mathcal{H}_{P}(y_1, y_2,\dots , y_m)$ by
    \begin{equation}
    	\mathcal{H}_P(y_1,y_2,\dots ,y_m)=\sum\limits_{v\in P}\prod_{i=1}^{m}y_i^{v_i}\in \mathbb{Z}[y_1,y_2,\dots ,y_m],
    \end{equation}
    where $v=(v_1,\dots ,v_m)$.\\
    We recall a lemma from \citep{Chang}.
    \begin{lemma}\citep{Chang}\label{generatinglemma}
    Suppose $\Delta\subset \mathbb{F}_{2}^m$ is a simplicial complex. If $\mathcal{F}$ is the collection of all maximal elements of $\Delta$, then 
    	\begin{equation}
    		\mathcal{H}_{\Delta}(y_1,y_2,\dots ,y_m)=\sum\limits_{\emptyset\neq S\subseteq\mathcal{F}}(-1)^{|S|+1}\prod_{i\in \cap S}(1+y_i),
    	\end{equation}
    where $\cap S=\bigcap\limits_{F\in S}\textnormal{Supp}(F)$. In particular, we have
    \begin{equation*}
    	|\Delta|=\sum\limits_{\emptyset\neq S\subseteq \mathcal{F}}(-1)^{|S|+1}2^{|\cap S|}.
    \end{equation*}
    \end{lemma}
    \begin{example}
    	Consider the simplicial complex $$\Delta=\{(0,0,0,0), (1,0,0,0), (0,1,0,0), (0,0,0,1), (1,1,0,0), (0,1,0,1)\}\subset \mathbb{F}_{2}^4.$$
    	Then $\mathcal{F}=\{F_1, F_2\}$, where $F_1=(1,1,0,0), F_2=(0,1,0,1)$. So
    	\begin{equation*}
    		\begin{split}
    			\mathcal{H}_{\Delta}(y_1,y_2,y_3,y_4)=&\prod_{i\in \textnormal{Supp}(F_1)}(1+y_i)+\prod_{i\in \textnormal{Supp}(F_2)}(1+y_i)-(1+y_2)\\=&(1+y_1)(1+y_2)+(1+y_2)(1+y_4)-(1+y_2)\\
    			=&1+y_1+y_2+y_4+y_1y_2+y_2y_4
    		\end{split}
    	\end{equation*}
    and $|\Delta|=6.$
    \end{example}    
    \subsection{Construction of subfield codes}
    Let $\mathbb{F}_{8}=\mathbb{F}_2(\omega)$, where $\omega \in \mathbb{F}_{8}$ satisfies the polynomial $y^3+y+1=0$. Now for each $v\in \mathbb{F}_{8}^m$, where $m\in \mathbb{N}$, there exist unique $a,b,c\in \mathbb{F}_{2}^m$ such that $v=a+\omega b+\omega^2 c$. Suppose that $C$ is a $k$-dimensional code of length $n$ over $\mathbb{F}_{q^m}$ generated by $G$ and $\mathcal{B}$ is an ordered basis of $\mathbb{F}_{q^m}$ over $\mathbb{F}_{q}$. The code $C^{(q)}$ over $\mathbb{F}_{q}$ generated by the matrix which is obtained by replacing each entry of $G$ by its column representation (a column vector in $\mathbb{F}_{q}^m$) with respect to $\mathcal{B}$, is termed as a subfield code.
    Now we recall a result from \citep{Ding}.
    \begin{lemma}\citep{Ding}\label{matrixlemma}
    	Suppose $\mathcal{B}=\{v_1<v_2<\dots <v_m\}$ is an ordered basis of $\mathbb{F}_{q^m}$ over $\mathbb{F}_{q}$. Suppose that $C$ is a $k$-dimensional linear code of length $n$ over $\mathbb{F}_{q^m}$ generated by
    	\begin{equation*}
    		G=\begin{pmatrix}
    			g_{11}&g_{12}&\dots&g_{1n}\\
    			g_{21}&g_{22}&\dots&g_{2n}\\
    			\vdots&\vdots&\ddots&\vdots\\
    			g_{k1}&g_{k2}&\dots&g_{kn}
    		\end{pmatrix}.
    	\end{equation*}
     Then the subfield code $C^{(q)}$ is a binary linear code generated by
     \begin{equation*}
     	G^{(q)}=\begin{pmatrix}
     		G_{1}^{(q)}\\
     		G_{2}^{(q)}\\
     		\vdots\\
     		G_{k}^{(q)}
     	\end{pmatrix},
     \end{equation*}
     where $Tr_{q^m/q}: \mathbb{F}_{q^m}\longrightarrow \mathbb{F}_{q}$ is the trace map and, for $1\leq i \leq k$
     \begin{equation*}
     	G_{i}^{(q)}=\begin{pmatrix}
     		Tr_{q^m/q}(g_{i1}v_1)&Tr_{q^m/q}(g_{i2}v_1)&\dots&Tr_{q^m/q}(g_{in}v_1)\\
     		Tr_{q^m/q}(g_{i1}v_2)&Tr_{q^m/q}(g_{i2}v_2)&\dots&Tr_{q^m/q}(g_{in}v_2)\\
     		\vdots&\vdots&\ddots&\vdots\\
     		Tr_{q^m/q}(g_{i1}v_m)&Tr_{q^m/q}(g_{i2}v_m)&\dots&Tr_{q^m/q}(g_{in}v_m)
     	\end{pmatrix}.
     \end{equation*}
     \end{lemma}
     Set $m=3, q=2$ and fix the basis $\{v_1=1, v_2=\omega, v_3=\omega^2\}$ for $\mathbb{F}_8$ over $\mathbb{F}_2$ in Lemma \ref{matrixlemma}. Let $g_{ij}=g_{ij}^{(1)}+\omega g_{ij}^{(2)}+\omega^2 g_{ij}^{(3)}\in \mathbb{F}_8$, where $g_{ij}^{(1)}, g_{ij}^{(2)}, g_{ij}^{(3)}\in \mathbb{F}_{2}$. Then we have, $Tr_{8/2}(g_{ij}v_1)=g_{ij}^{(1)}$, $Tr_{8/2}(g_{ij}v_2)=g_{ij}^{(3)}$ and $Tr_{8/2}(g_{ij}v_3)=g_{ij}^{(2)}$. Now, we have the following result.
     
     \begin{theorem}\label{matrixtheorem}
     Let $\mathcal{B}=\{1< \omega < \omega^2\}$ be the ordered basis of $\mathbb{F}_{8}$ over $\mathbb{F}_{2}$. Suppose $G:=G_1+\omega G_2+\omega^2G_3\in M_{k\times n}(\mathbb{F}_{8})$ generates the linear code $C$ over $\mathbb{F}_{8}$, where $G_i\in M_{k\times n}(\mathbb{F}_{2})$, $1\leq i\leq 3$. Then the code $C^{(2)}$, with respect to $\mathcal{B}$, is a linear code over $\mathbb{F}_2$ generated by
      \begin{equation*}
      	G^{(2)}=\begin{pmatrix}
      		G_1\\
      		G_3\\
      		G_2
      	\end{pmatrix}.
      \end{equation*}
      In addition, $C_{D}^{(2)}=C_{D^{(2)}}$ that is, if the defining set of the code $C_{D}$ is $D=D_1+\omega D_2+\omega^2D_3\subset \mathbb{F}_{8}^m$, where $D_i \subset \mathbb{F}_{2}^m, 1\leq i\leq 3$, then the defining set of $C_{D}^{(2)}$ is
      \begin{equation*}
      	D^{(2)}=\{(d_1,d_3,d_2)| d_i\in D_i, 1\leq i\leq 3\}.
      \end{equation*}
      \begin{proof}
      	Proof follows from the above discussion.
      \end{proof}
      \end{theorem}
      The Theorem \ref{matrixtheorem} is illustrated by the following example.
      \begin{example}
      	Consider the ordered basis $\{1< \omega < \omega^2\}$ of $\mathbb{F}_{8}$ over $\mathbb{F}_{2}$.
      	Let $D_1=\{(1,0)\}, D_2=\{(0,1), (1,1)\}$ and $D_3=\{(1,0), (0,1), (1,1)\}$ so that $D=D_1+\omega D_2+\omega^2D_3\subset \mathbb{F}_{8}^2$ is the defining set of the code $C_{D}$. Consider the following generator matrix of $C_{D}$:
      	\begin{equation*}
      		\begin{split}
      		G=& \begin{pmatrix}
      			1 & 1 & \omega^2 & \omega^2 & 1+\omega+\omega^2 & 1+\omega+\omega^2\\
      			0 & 1 & 1 & \omega^2 & 1 & 1+\omega
      		\end{pmatrix}\\
      		=&\begin{pmatrix}
      			1 & 1 & 0 & 0 & 1 & 1\\
      			0 & 1 & 1 & 0 & 1 & 1
      		\end{pmatrix}
      		+\omega \begin{pmatrix}
      			0 & 0 & 0 & 0 & 1 & 1\\
      			0 & 0 & 0 & 0 & 0 & 1
      		\end{pmatrix}
      		+\omega^2 \begin{pmatrix}
      			0 & 0 & 1 & 1 & 1 & 1\\
      			0 & 0 & 0 & 1 & 0 & 0
      		\end{pmatrix}.
      	    \end{split}
      	\end{equation*}
      Then $C_{D}$ is a $[6, 2, 4]$-linear code over $\mathbb{F}_{8}$ and according to Theorem \ref{matrixtheorem}, a generator matrix and a defining set of $C^{(2)}_{D}$ are given by 
       \begin{equation*}      	
      	G^{(2)}=\begin{pmatrix}
      		1 & 1 & 0 & 0 & 1 & 1\\
      		0 & 1 & 1 & 0 & 1 & 1\\
      		0 & 0 & 1 & 1 & 1 & 1\\
      		0 & 0 & 0 & 1 & 0 & 0\\
      		0 & 0 & 0 & 0 & 1 & 1\\
      		0 & 0 & 0 & 0 & 0 & 1
      	\end{pmatrix}   	
      \end{equation*}
     and $D^{(2)}=\{(1,0,0,1,1,0), (1,0,0,1,0,1), (1,0,0,1,1,1), (1,0,1,1,1,0), (1,0,1,1,0,1), (1,0,1,1,1,1)\}$, respectively.
      \end{example}
      \section{Linear codes using simplicial complexes}\label{section_3}
       This section studies linear codes over $\mathbb{F}_{2^3}$ and its generalization to the field $\mathbb{F}_{2^n}$  with the help of simplicial complexes.\par
       Suppose that $\mathbb{F}_{8}=\mathbb{F}_{2}(\omega)$, where $\omega \in \mathbb{F}_{8}$ satisfying $y^3+y+1=0$. Let $m\in \mathbb{N}$ and let $D_i\subseteq \mathbb{F}_{2}^m, 1\leq i \leq 3$. Assume that $D=D_1+\omega D_2+\omega^2D_3 \subseteq \mathbb{F}_{8}^m$. Define an octanary linear code as follows:
       \begin{equation}\label{constructionofCD}
       	C_{D^{\ast}}=\{(v\cdot d)_{d\in D^{\ast}}| v\in \mathbb{F}_{8}^m\}.
       \end{equation}
       Observe that the map $c_{D^{\ast}}: \mathbb{F}^m_{2^3}\longrightarrow C_{D^{\ast}}$ defined by $c_{D^{\ast}}(v)=(v\cdot d)_{d\in D^{\ast}}$ is a surjective linear transformation. By definition, $|D^{\ast}|$ is the length of $C_{D^{\ast}}$. Assume that $v=\alpha +\omega \beta +\omega^2\gamma \in \mathbb{F}_{8}^m$ and $d=d_1+\omega d_2+\omega^2 d_3$, where $\alpha, \beta, \gamma \in \mathbb{F}_{2}^m$ and $d_i\in D_i, 1\leq i\leq 3$. Then we have
       \begin{equation*}
       	\begin{split}
       		wt(c_{D^{\ast}}(v)) = & wt\big(((\alpha+\omega \beta +\omega^2\gamma)\cdot(d_1+\omega d_2+\omega^2d_3))_{d_1\in D_1, d_2\in D_2, d_3\in D_3}\big)\\
       		= & wt\big((\alpha \cdot d_1+\gamma \cdot d_2+\beta \cdot d_3)+\omega\big(\beta \cdot d_1+(\alpha+\gamma)\cdot d_2+(\beta+\gamma)\cdot d_3\big)+\\
       		&~ \omega^2\big(\gamma \cdot d_1+\beta \cdot d_2+(\alpha +\gamma)\cdot d_3\big)_{d_1\in D_1, d_2\in D_2, d_3\in D_3}\big).
       	\end{split}
       \end{equation*}
      Now if $u=u_1 +\omega u_2+\omega^2 u_3\in \mathbb{F}_{8}^m$ with $u_1, u_2, u_3 \in \mathbb{F}_{2}^m$, then $wt(u)=0 \iff u_1=u_2=u_3=0$. Hence,
      \begin{equation*}
      	\begin{split}
      		wt(c_{D^{\ast}}(v)) =wt(c_{D}(v)) = & |D|-\frac{1}{8}\sum\limits_{d_1\in D_1}\sum\limits_{d_2\in D_2}\sum\limits_{d_3\in D_3}\big(1+(-1)^{\alpha\cdot  d_1+\gamma\cdot d_2 + \beta\cdot d_3} \big)\times \\
      		&~\big(1+(-1)^{\beta\cdot d_1+(\alpha+\gamma)\cdot d_2+(\beta+\gamma)\cdot d_3}\big)\times\big(1+(-1)^{\gamma\cdot d_1+\beta\cdot d_2+(\alpha+\gamma)\cdot d_3} \big).
      	\end{split}
      \end{equation*}
      Define for $x\in \mathbb{F}_2^m$,
      \begin{equation}
      	\chi_{x}: 2^{\mathbb{F}^m_{2}}\longrightarrow \mathbb{Z} \textnormal{  by  } \chi_{x}(P)=\sum\limits_{y\in P}(-1)^{x\cdot y}.
      \end{equation}
      Then,
      \begin{equation}\label{wtequation}
      	\begin{split}
      		wt(c_{D^{\ast}}(v))= &\frac{7}{8}|D|-\frac{1}{8}\big\{\chi_{\alpha}(D_1)\chi_{\gamma}(D_2)\chi_{\beta}(D_3)+\chi_{\beta}(D_1)\chi_{\alpha+\gamma}(D_2)\chi_{\beta+\gamma}(D_3)+\\
      		&~ \chi_{\gamma}(D_1)\chi_{\beta}(D_2)\chi_{\alpha+\gamma}(D_3) +\chi_{\alpha+\beta}(D_1)\chi_{\alpha}(D_2)\chi_{\gamma}(D_3)+\\
      		&~\chi_{\alpha+\gamma}(D_1)\chi_{\beta+\gamma}(D_2)\chi_{\alpha+\beta+\gamma}(D_3)+\chi_{\beta+\gamma}(D_1)\chi_{\alpha+\beta+\gamma}(D_2)\chi_{\alpha+\beta}(D_3)+\\
      		&~\chi_{\alpha+\beta+\gamma}(D_1)\chi_{\alpha+\beta}(D_2)\chi_{\alpha}(D_3)\big\}.
      	\end{split}
      \end{equation}
       Denoted by $\delta$ the Kronecker delta function and set $D^{c}=\mathbb{F}_{8}^m\setminus D$. Then 
       \begin{equation}\label{DcExpressionF8}
       		D^{c}=\big(D_1^{c}+\omega \mathbb{F}_{2}^m+\omega^2\mathbb{F}_{2}^m\big)\bigsqcup \big(D_1+\omega D_2^c+\omega^2\mathbb{F}_{2}^m\big)\bigsqcup \big(D_1+\omega D_2+\omega^2 D_3^c\big),
       \end{equation}
        where $\bigsqcup$ indicates disjoint union.\\
        By using the identity $\chi_{x}(\mathbb{F}_{2}^m)=2^{m}\delta_{0, x}$ and $\chi_{x}(P)+\chi_{x}(P^c)=\chi_{x}(\mathbb{F}_{2}^m)$ for $x\in \mathbb{F}_{2}^m, P\subset \mathbb{F}_{2}^m$, we get
        \begin{equation}
        	\begin{split}\label{wtequationforDc}
        		wt(c_{D^{c}}(v))=&\frac{7}{8}\big(|D^c|-2^{3m}\delta_{0,v}\big)+ \frac{1}{8}\big\{\chi_{\alpha}(D_1)\chi_{\gamma}(D_2)\chi_{\beta}(D_3)+\chi_{\beta}(D_1)\chi_{\alpha+\gamma}(D_2)\chi_{\beta+\gamma}(D_3)+\\
        		&~ \chi_{\gamma}(D_1)\chi_{\beta}(D_2)\chi_{\alpha+\gamma}(D_3) +\chi_{\alpha+\beta}(D_1)\chi_{\alpha}(D_2)\chi_{\gamma}(D_3)+\\
        		&~\chi_{\alpha+\gamma}(D_1)\chi_{\beta+\gamma}(D_2)\chi_{\alpha+\beta+\gamma}(D_3)+\chi_{\beta+\gamma}(D_1)\chi_{\alpha+\beta+\gamma}(D_2)\chi_{\alpha+\beta}(D_3)+\\
        		&~\chi_{\alpha+\beta+\gamma}(D_1)\chi_{\alpha+\beta}(D_2)\chi_{\alpha}(D_3)\big\}.
        	\end{split}
        \end{equation}
       By Equation \ref{wtequation} and \ref{wtequationforDc}, we have the following lemma.
       \begin{lemma}\label{wtequationrelation1}
       	Let the symbols be same as above. For $v\in \mathbb{F}_{8}^m$, we have
       	\begin{equation}
       		\begin{split}
       			wt(c_{D^c}(v))+ wt(c_{D^{\ast}}(v))= 7\times 2^{3(m-1)}\big(1-\delta_{0,v}\big).
       		\end{split}
       	\end{equation}
       \end{lemma}
       
      By a similar fashion, we can generalize the above relation for codes over $\mathbb{F}_{2^n}$.\\
      Let $\mathbb{F}_{2^n}=\mathbb{F}_{2}(\eta)$ be the finite field of order $2^n$, where $\eta \in \mathbb{F}_{2^n}$. Consider the defining set $D=D_1+\eta D_2+\cdots +\eta^{n-1}D_n\subset \mathbb{F}_{2^n}^m$, where $m\in \mathbb{N}$ and $D_i\subset \mathbb{F}_{2}^m$ for $1\leq i\leq n$. Then
   \begin{equation}\label{generalDc}
   	\begin{split}
   		D^{c} = &\big(D_1^c+\eta\mathbb{F}_{2}^m+\cdots +\eta^{n-1}\mathbb{F}_{2}^m\big)\bigsqcup\\& \big(D_1+\eta D_2^c+\eta^2\mathbb{F}_{2}^m\cdots +\eta^{n-1}\mathbb{F}_{2}^m\big)\bigsqcup \\&
   		~\vdots\\& 
   		\big(D_1+\eta D_2+\cdots +\eta^{n-2} D_{n-1}^c +\eta^{n-1}\mathbb{F}_{2}^m\big)\bigsqcup \\ 
   		& \big(D_1+\eta D_2+\cdots +\eta^{n-2} D_{n-1} +\eta^{n-1}D_n^c\big).
   	\end{split}
   \end{equation}
    Now we conjecture that the relation in Lemma \ref{wtequationrelation1} get extended for codes over any finite field of characteristic $2$.
    \begin{conjecture}
    	Let the symbols be same as above. For $v\in \mathbb{F}_{2^n}^m$, we have
    	\begin{equation}\label{generalwtrelation}
    		\begin{split}
    			wt(c_{D^c}(v)) + wt(c_{D^{\ast}}(v))=& (2^{n}-1)\times 2^{n(m-1)}\big(1-\delta_{0,v}\big).
    		\end{split}
    	\end{equation}
    \end{conjecture}
    In the next result we find the parameters of a linear code over $\mathbb{F}_{8}$.
    \begin{proposition}\label{DtheoremF8}
    	Consider the defining set $D=\Delta_{L}+\omega \Delta_{M}+\omega^2\Delta_{N}\subset \mathbb{F}_{8}^m$, where $m\in \mathbb{N}$ and $L, M, N$ are nonempty subsets of $[m]$. Assume that at least two of the sets $L\setminus (M\cup N), M\setminus (N\cup L),$ and $N\setminus (L\cup M)$ are nonempty. Then the code $C_{D^{\ast}}$ is a $3$-weight linear code over $\mathbb{F}_{8}$ of length $2^{|L|+|M|+|N|}-1$, dimension $|L\cup M\cup N|$ and distance $2^{|L|+|M|+|N|-1}$. In particular, $C_{D^{\ast}}$ have codewords of weights $0, 2^{|L|+|M|+|N|-1}, 6\times 2^{|L|+|M|+|N|-3}$ and $7\times 2^{|L|+|M|+|N|-3}$. If $(1, A_1, A_2,\dots, A_{|D^{\ast}|})$ is the weight distribution of $C_{D^{\ast}}$ and  $Z_{i} =|\{v\in \mathbb{F}_{2^3}^m : wt(c_{D^{\ast}}(v))=i \}|$ for $0\leq i\leq |D^{\ast}|$, then
    	\begin{equation*}       		
    		Z_{0}=2^{3(m-|L\cup M\cup N|)} \textnormal{ and } Z_i=Z_0A_i.    		
    	\end{equation*}
    	\begin{proof}
    		Observe the length of the code $C_{D^{\ast}}$ is $2^{|L|+|M|+|N|}-1$. Consider the map $c_{D^{\ast}} : \mathbb{F}_{2^3}^m \longrightarrow C_{D^{\ast}}$ defined by $c_{D^{\ast}}(x)=(x\cdot d)_{d\in D^{\ast}}$ which is a surjective linear transformation and
    		\begin{equation}\label{KerOfc_DF2n}
    			\begin{split}
    				\ker(c_{D^{\ast}})=&\{v=(v_i)\in \mathbb{F}_{2^3}^m: v\cdot d=0~\forall ~d\in D^{\ast}\}\\
    				=&\{v=(v_i)\in \mathbb{F}_{2^3}^m: v_i=0 \textnormal{ for }i\in L \cup M\cup N\}.
    			\end{split}
    		\end{equation}
    		Therefore, $|\ker(c_{D^{\ast}})|=2^{3({m-|L\cup M\cup N|})}=Z_{0}$. By the first isomorphism theorem of groups, we have $|C_{D^{\ast}}|=\frac{|\mathbb{F}_{2^3}^m|}{|\ker(c_{D^{\ast}})|}=2^{3|L\cup M \cup N|}$. Hence $\dim (C_{D^{\ast}})=|L\cup M\cup N|$.\\
    		Note that $\textnormal{Supp}(d)\subseteq L\cup M \cup N$ for $d\in D$. Let $Y_i=\{c\in C_{D^{\ast}}: wt(c)=i\}$ so that $|Y_i|=A_{i}$ and let $X_{i}=\{v\in \mathbb{F}_{2^3}^m: wt(c_{D^{\ast}}(v))=i\}$ so that $|X_i|=Z_i$ for $0\leq i\leq |D^{\ast}|$. Since $Y_i=c_{D^{\ast}}(X_i)$, we have $X_i=c_{D^{\ast}}^{-1}(Y_i)= \displaystyle{\bigsqcup_{c \in Y_i} c^{-1}_{D^{\ast}}(c)}$. Let $c\in C_{D^{\ast}}$ then $|c^{-1}_{D^{\ast}}(c)|=|c^{-1}_{D^{\ast}}(0)|=|\ker(c_{D^{\ast}})|=Z_0$. Therefore $Z_i=|X_i|=\sum\limits_{c\in Y_i}|c^{-1}_{D^{\ast}}(c)|=Z_0A_i$.\\    		
    		For $Y\subseteq [m]$, define the Boolean function $\varphi(\cdot|Y): \mathbb{F}_{2}^m\longrightarrow \mathbb{F}_{2}$ as
    		\begin{equation*}
    			\varphi(x|Y)=\prod_{i\in Y}(1-x_i)=\begin{cases}
    				1, \text{  if } \textnormal{ Supp}(x)\cap Y=\emptyset;\\
    				0, \text{  if } \textnormal{ Supp}(x)\cap Y\neq \emptyset.
    			\end{cases}
    		\end{equation*}
    		For $x=(x_1,\dots ,x_m)\in \mathbb{F}_{2}^m$ and $\emptyset \neq L\subset [m]$, we have
    		\begin{equation}\label{chiequation1}
    			\begin{split}
    				\chi_{x}(\Delta_{L})=&\sum\limits_{y\in \Delta_{L}}(-1)^{x\cdot y}  =\mathcal{H}_{\Delta_{L}}((-1)^{x_1},\dots ,(-1)^{x_m})\\
    				=&\prod_{i\in L}\big(1+(-1)^{x_i}\big)
    				=\prod_{i\in L}(2-2x_i)\\=&2^{|L|}\prod_{i\in L}(1-x_i)=2^{|L|}\varphi(x|L).
    			\end{split}
    		\end{equation}
    		Let $v=\alpha+\omega \beta+\omega^2\gamma\in \mathbb{F}_{8}^m$. By Equation \ref{wtequation} and \ref{chiequation1}, we have
    		\begin{equation*}
    			\begin{split}
    				wt(c_{D^{\ast}}(v))=& wt(c_{D}(v))\\
    				=&\frac{7}{8}|D|-2^{|L|+|M|+|N|-3}\big\{\varphi(\alpha|L)\varphi(\gamma|M)\varphi(\beta|N)+\varphi(\beta|L)\varphi(\alpha+\gamma|M)\varphi(\beta+\gamma|N)+\\&\varphi(\gamma|L)\varphi(\beta|M)\varphi(\alpha+\gamma|N)+\varphi(\alpha+\beta|L)\varphi(\alpha|M)\varphi(\gamma|N)+\\&\varphi(\alpha+\gamma|L)\varphi(\beta+\gamma|M)\varphi(\alpha+\beta+\gamma|N)+\varphi(\beta+\gamma|L)\varphi(\alpha+\beta+\gamma|M)\varphi(\alpha+\beta|N)+\\&\varphi(\alpha+\beta+\gamma|L)\varphi(\alpha+\beta|M)\varphi(\alpha|N)\big\}.
    			\end{split}
    		\end{equation*}
    		Let
    		\begin{equation}\label{thetaEquation}
    			\begin{split}
    				\theta= &\varphi(\alpha|L)\varphi(\gamma|M)\varphi(\beta|N)+\varphi(\beta|L)\varphi(\alpha+\gamma|M)\varphi(\beta+\gamma|N)+ \varphi(\gamma|L)\varphi(\beta|M)\varphi(\alpha+\gamma|N)+\\
    				&\varphi(\alpha+\beta|L)\varphi(\alpha|M)\varphi(\gamma|N)+ \varphi(\alpha+\gamma|L)\varphi(\beta+\gamma|M)\varphi(\alpha+\beta+\gamma|N)+\\
    				&\varphi(\beta+\gamma|L)\varphi(\alpha+\beta+\gamma|M)\varphi(\alpha+\beta|N)+ \varphi(\alpha+\beta+\gamma|L)\varphi(\alpha+\beta|M)\varphi(\alpha|N).
    			\end{split}   
    		\end{equation}
    		\textnormal{Case: }$\theta=7$\\
    		If $\theta=7$ then $wt(c_{D^{\ast}}(v))=0$ and $\theta=7 \iff \varphi(\alpha|L)=\varphi(\beta|L)=\varphi(\gamma|L)=\varphi(\alpha|M)=\varphi(\beta|M)=\varphi(\gamma|M)=\varphi(\alpha|N)=\varphi(\beta|N)=\varphi(\gamma|N)=1$.\\ Therefore,
    		$\textnormal{Supp}(\alpha)\cap (L\cup M\cup N)=\textnormal{Supp}(\beta)\cap (L\cup M\cup N)=\textnormal{Supp}(\gamma)\cap (L\cup M\cup N)=\emptyset$.\\
    		Hence, the number of such $v=\alpha +\omega \beta+\omega^2\gamma$ is $8^{m-|L\cup M\cup N|}$ and an example of such $v$ for which $\theta=7$ is  $v=(0,0,\dots,0)$.\\
    		\textnormal{Case: }$\theta=6$\\
    		Let $\theta=6$. Then the Equation \ref{thetaEquation} holds if and only if exactly one of the term in it is $0$ and rest are all equal to $1$. For example
    		\begin{equation*}
    			\begin{split}
    				\varphi(\alpha|L)\varphi(\gamma|M)\varphi(\beta|N)&=0,\\
    				\varphi(\beta|L)\varphi(\alpha+\gamma|M)\varphi(\beta+\gamma|N)&=1,\\ \varphi(\gamma|L)\varphi(\beta|M)\varphi(\alpha+\gamma|N)&=1,\\ \varphi(\alpha+\beta|L)\varphi(\alpha|M)\varphi(\gamma|N)&=1,\\ \varphi(\alpha+\gamma|L)\varphi(\beta+\gamma|M)\varphi(\alpha+\beta+\gamma|N)&=1,\\ \varphi(\beta+\gamma|L)\varphi(\alpha+\beta+\gamma|M)\varphi(\alpha+\beta|N)&=1,\\ \varphi(\alpha+\beta+\gamma|L)\varphi(\alpha+\beta|M)\varphi(\alpha|N)&=1.
    			\end{split}
    		\end{equation*}
    	   One can verify that these equations can not hold simultaneously. Similarly other six possibilities are also not possible. So that the case $\theta=6$ is rejected. The same argument can be used to reject the possibilities of $\theta=5, 4$ and $2$.\\
    		\textnormal{Case: }$\theta=3$\\
    		If $\theta=3$ then we have
    		\begin{equation*}
    			\begin{split}
    				wt(c_{D^{\ast}}(v))=&\frac{7}{8}|D|-\frac{3}{8}2^{|L|+|M|+|N|}\\
    				=& 2^{|L|+|M|+|N|-1}
    			\end{split}
    		\end{equation*}
    		and an example of such $v$ for which $\theta=3$ is $v=\alpha$ with $\textnormal{Supp}(\alpha)=\{i\}$ where $i\in L\setminus (M\cup N)$.\\
    		\textnormal{Case: }$\theta=1$\\
    		If $\theta=1$ then we have
    		\begin{equation*}
    			\begin{split}
    				wt(c_{D^{\ast}}(v))=&\frac{7}{8}|D|-\frac{1}{8}2^{|L|+|M|+|N|}\\
    				=&6\times 2^{|L|+|M|+|N|-3}
    			\end{split}
    		\end{equation*}
    		and an example of such $v$ for which $\theta=1$ is $v=\alpha$ with $\textnormal{Supp}(\alpha)=\{i,j\}$ where $i\in L\setminus (M\cup N), j\in M\setminus (N\cup L)$.\\ 
    		\textnormal{Case: }$\theta=0$\\
    		If $\theta=0$ then we have $wt(c_{D^{\ast}}(v))=7\times 2^{|L|+|M|+|N|-3}$ and an example of such $v$ for which $\theta=0$ is $v=\alpha=(1,1,\dots,1)$ where $\textnormal{Supp}(\alpha)\cap L, \textnormal{Supp}(\alpha)\cap M, \textnormal{Supp}(\alpha)\cap N$ are each nonempty.    		
    	\end{proof}
    \end{proposition}
     If all the three subsets $L, M$ and $N$ are nonempty and equal then we have the following.
    \begin{theorem}\label{CorCDF8}
    	Let $\emptyset \neq L=M=N\subsetneq [m]$ and let the defining set $D=\Delta_{L}+\omega \Delta_{L}+\omega^2\Delta_{L}\subset \mathbb{F}_{8}^m$, where $m\in \mathbb{N}$. Then the code $C_{D^{\ast}}$ is a $1$-weight linear code over $\mathbb{F}_{8}$ of length $2^{3|L|}-1$, dimension $|L|$ and distance $7\times 2^{3(|L|-1)}$. In particular, it is a minimal code. Moreover, $C_{D^{\ast}}$ is a Griesmer code and hence it is a distance optimal. If $(1, A_1, A_2,\dots, A_{|D^{\ast}|})$ is the weight distribution of $C_{D^{\ast}}$ and $Z_{i} =|\{v\in \mathbb{F}_{2^3}^m : wt(c_{D^{\ast}}(v))=i \}|$ for $0\leq i\leq |D^{\ast}|$, then
    	\begin{equation*}       		
    		Z_{0}=2^{3(m-|L|)} \textnormal{ and } Z_i=Z_0A_i.    		
    	\end{equation*}
    \end{theorem}
    Proposition \ref{DtheoremF8} and Theorem \ref{CorCDF8} are illustrated in the following examples.
    \begin{example}\label{multiplicityexample}
    	Set $m=4$ and $L=\{1, 2\}, M=\{2,3\}, N=\{2\}\subseteq [4]=\{1, 2, 3, 4\}$ and consider the defining set $D=\Delta_{L}+\omega \Delta_{M}+\omega^2\Delta_{N}\subset \mathbb{F}_{8}^3$, where
    	\begin{equation*}
    		\begin{split}
    			\Delta_{L}=&\{(0,0,0,0), (1,0,0,0), (0, 1, 0,0), (1, 1, 0,0)\},\\ \Delta_{M}=&\{(0,0,0,0), (0,1,0,0), (0,0,1,0), (0,1,1,0)\},\\
    			\Delta_{N}=&\{(0,0,0,0), (0,1,0,0)\}.
    		\end{split}
    	\end{equation*}
    Note that $C_{D^{\ast}}$ is a $[31, 3, 16]$ linear code over $\mathbb{F}_{8}$ which is not optimal; in fact an optimal octanary linear code of length $31$ and dimension $3$ has distance $26$ (see \citep{BoundTable}).
    \end{example}
    \begin{example}
    	Set $m=3$ and $L=\{1, 2\}= M= N$. Let $D=\Delta_{L}+\omega \Delta_{L}+\omega^2\Delta_{L}\subset \mathbb{F}_{8}^2$, where
    	\begin{equation*}
    		\begin{split}
    			\Delta_{L}=&\{(0,0,0), (1,0,0), (0,1,0), (1,1,0)\}.
    		\end{split}
    	\end{equation*}
    	Note that $C_{D^{\ast}}$ is a $[63, 2, 56]$ linear code over $\mathbb{F}_{8}$ which is distance optimal  from \citep{BoundTable}.
    \end{example}
      The following result gives parameters of codes over the field $\mathbb{F}_{2^n}$.
    \begin{proposition}\label{theoremoverfield2n}
       	Let $\mathbb{F}_{2^n}=\mathbb{F}_{2}(\eta)$, where $\eta \in \mathbb{F}_{2^n}$. Consider the defining set $D=\Delta_{M_1}+\eta \Delta_{M_2}+\cdots +\eta^{n-1}\Delta_{M_n}\subset \mathbb{F}_{2^n}^m$, where $m\in \mathbb{N}$ and $\emptyset \neq M_{i}\subsetneq [m], 1\leq i\leq n$. Then the code $C_{D^{\ast}}$ is a $[2^{\sum_{j=1}^n|M_{j}|}-1, |\cup_{j=1}^{n}M_j|]$ linear code over $\mathbb{F}_{2^n}$.	If $(1, A_1, A_2,\dots, A_{|D^{\ast}|})$ is the weight distribution of $C_{D^{\ast}}$ and $Z_{i} =|\{v\in \mathbb{F}_{2^n}^m : wt(c_{D^{\ast}}(v))=i \}|$, where $0\leq i\leq |D^{\ast}|$, then
       	\begin{equation*}       		
       			Z_{0}=2^{n(m-|\cup_{j=1}^{n}M_j|)} \textnormal{ and } Z_i=Z_0A_i\textnormal{  where  }1\leq i\leq |D^{\ast}|.
       	\end{equation*}       
       \begin{proof}       
       	Proof follows from the argument used in the proof of Proposition \ref{DtheoremF8}.
       \end{proof}
       \end{proposition}
        Based on the computations done for codes over the fields $\mathbb{F}_{2^r}, r=1, 2, 3$ and $4$, we conjecture the following.
       \begin{conjecture}
       	Let $\emptyset \neq M_{i}=M\subsetneq [m], 1\leq i\leq n$ and let the defining set
       	$D=\Delta_{M}+\eta \Delta_{M}+\cdots +\eta^{n-1}\Delta_{M}\subset \mathbb{F}_{2^n}^m$, where $m\in \mathbb{N}$. Then the code $C_{D^{\ast}}$ is a $1$-weight linear code over $\mathbb{F}_{2^n}$ of length $2^{n|M|}-1$, dimension $|M|$ and distance $(2^n-1)\times 2^{n(|M|-1)}$. In particular, it is a minimal code. Moreover, $C_{D^{\ast}}$ is a Griesmer code. If $(1, A_1, A_2,\dots, A_{|D^{\ast}|})$ is the weight distribution of $C_{D^{\ast}}$ and $Z_{i} =|\{v\in \mathbb{F}_{2^n}^m : wt(c_{D^{\ast}}(v))=i \}|$, where $0\leq i\leq |D^{\ast}|$, then
       	\begin{equation*}       		
       		Z_{0}=2^{n(m-|M|)} \textnormal{ and } Z_i=Z_0A_i.
       	\end{equation*}
       \end{conjecture}
              
       In the next result, we find the parameters of a code whose defining set is the complement of the defining set in Proposition \ref{DtheoremF8}.
   \begin{theorem}\label{theoremF8Dc}
   	Let $m\in \mathbb{N}$. Suppose that $L, M, N$ are nonempty subsets of $[m]$ such that $L\cup M\cup N\subsetneq [m]$. Further, assume that at least two of the sets $L\setminus (M\cup N), M\setminus (N\cup L),$ and $N\setminus (L\cup M)$ are nonempty. Let $D=\Delta_{L}+\omega \Delta_{M}+\omega^2\Delta_{N}\subset \mathbb{F}_{8}^m$ so that $D^{c}=\big(\Delta_{L}^{c}+\omega \mathbb{F}_{2}^m+\omega^2\mathbb{F}_{2}^m\big)\bigsqcup \big(\Delta_{L}+\omega \Delta_{M}^c+\omega^2\mathbb{F}_{2}^m\big)\bigsqcup \big(\Delta_{L}+\omega \Delta_{M}+\omega^2 \Delta_{N}^c\big)$. Then $C_{D^{c}}$ is a $[2^{3m}-2^{|L|+|M|+|N|}, m, 7\times 2^{3(m-1)}-7\times 2^{|L|+|M|+|N|-3}]$ octanary $4$-weight linear code. In particular, $C_{D^c}$ have codewords of weights $0, 7\times 2^{3(m-1)}-7\times 2^{|L|+|M|+|N|-3}, 7\times 2^{3(m-1)}-6\times 2^{|L|+|M|+|N|-3}, 7\times 2^{3(m-1)}-4\times 2^{|L|+|M|+|N|-3}$ and $7\times 2^{3(m-1)}$. Moreover, $C_{D^{c}}$ is a Griesmer code and hence it is a distance optimal. In fact, it is a minimal code if $|L|+|M|+|N|\leq 3m-4$.
   	\begin{proof}
   		The code $C_{D^c}$ is a $4$-weight octanary linear code with the required parameters by using Lemma \ref{wtequationrelation1}.\\
   		Now we have
   		\begin{equation*}
   			\begin{split}
   				\sum\limits_{i=0}^{m-1}\left\lceil \frac{d}{q^i}\right\rceil=&
   				\sum\limits_{i=0}^{m-1}\left\lceil \frac{7\times 2^{3m-3}-7\times 2^{|L|+|M|+|N|-3}}{8^i}\right\rceil\\ =&\sum\limits_{i=0}^{m-1} \frac{7\times 2^{3m-3}}{8^i}-\sum\limits_{i=0}^{m-1}\left\lfloor \frac{7\times 2^{|L|+|M|+|N|-3}}{8^i}\right\rfloor\\
   				= &7\times \big(2^{3m-3}+2^{3m-6}+\cdots +1\big)-\\ &  \big(7\times 2^{|L|+|M|+|N|-3}+7\times 2^{|L|+|M|+|N|-6}+\cdots +X+Y\big)
   			\end{split}
   		\end{equation*}
   		where,\\ if $|L|+|M|+|N|-3 \equiv 0\textnormal{ mod }3$ then $X=7, Y=0$;\\
   		if $|L|+|M|+|N|-3\equiv 1 \textnormal{ mod }3$ then $X=7\times 2, Y=1$;\\
   		if $|L|+|M|+|N|-3\equiv 2 \textnormal{ mod }3$ then $X=7\times 2^2, Y=3$.\\   		
   		Then
   		\begin{equation*}
   			\begin{split}
   				\sum\limits_{i=0}^{m-1}\left\lceil \frac{d}{q^i}\right\rceil =&\big(2^{3m }-1\big)-\big(2^{|L|+|M|+|N|}-1\big)\\
   				=& 2^{3m}-2^{|L|+|M|+|N|}.
   			\end{split}
   		\end{equation*}
   		Hence $C_{D^{c}}$ is a Griesmer code. By Lemma \ref{minimal_lemma}, we have
   		\begin{equation*}
   			\begin{split}
   				\frac{wt_{min}}{wt_{max}} =&\frac{7\times 2^{3m-3}-7\times 2^{|L|+|M|+|N|-3}}{7\times 2^{3m-3}}\\
   				=& 1-2^{|L|+|M|+|N|-3m}
   			\end{split}
   		\end{equation*}
   	  and $1-2^{|L|+|M|+|N|-3m}>\frac{7}{8} \iff |L|+|M|+|N|\leq 3m-4$.\\
   	  Hence, $C_{D^c}$ is minimal if $|L|+|M|+|N|\leq 3m-4$.
   	\end{proof}
   \end{theorem}
   If all the three subsets $L, M$ and $N$ of $[m]$ are nonempty and equal then we have the following.
   \begin{theorem}\label{CorCDcF8}
   	Let $\emptyset \neq L=M=N\subsetneq [m]$ and let the defining set $D=\Delta_{L}+\omega \Delta_{L}+\omega^2\Delta_{L}\subset \mathbb{F}_{8}^m$, where $m\in \mathbb{N}$. The code $C_{D^{c}}$ is a $2$-weight linear code over $\mathbb{F}_{8}$ of length $2^{3m}-2^{3|L|}$, dimension $m$ and distance $7\times 2^{3(m-1)}-7\times 2^{3(|L|-1)}$. In particular, $C_{D^c}$ have codewords of weights $0, 7\times 2^{3(m-1)}-7\times 2^{3(|L|-1)}$ and $7\times 2^{3(m-1)}$. Moreover, $C_{D^{c}}$ is a Griesmer code and hence it is a distance optimal. Further, it is a minimal code if $3(m-|L|)-4\geq 0$.
   \end{theorem}

   Here are examples illustrating Theorem \ref{theoremF8Dc} and Theorem \ref{CorCDcF8}.
   \begin{example}
   	Set $m=4$ and $L=\{1, 2\}, M=\{2,3\}, N=\{2\}\subseteq [4]=\{1, 2, 3, 4\}$. Let the defining set $D=\Delta_{L}+\omega \Delta_{M}+\omega^2\Delta_{N}\subset \mathbb{F}_{8}^4$ so that $D^{c}=\big(\Delta_{L}^{c}+\omega \mathbb{F}_{2}^4+\omega^2\mathbb{F}_{2}^4\big)\bigsqcup \big(\Delta_{L}+\omega \Delta_{M}^c+\omega^2\mathbb{F}_{2}^4\big)\bigsqcup \big(\Delta_{L}+\omega \Delta_{M}+\omega^2 \Delta_{N}^c\big)$, where
   	\begin{equation*}
   		\begin{split}
   			\Delta^{c}_{L}=&\{(0,0,1,0), (1,0,1,0), (0, 1, 1,0), (1, 1, 1,0), (0,0,0,1), (1,0,0,1), (0, 1, 0,1), (1, 1, 0,1), \\
   			&~ (0,0,1,1), (1,0,1,1), (0, 1, 1,1), (1, 1, 1,1)\},\\ \Delta^{c}_{M}=&\{(1,0,0,0), (1,1,0,0), (1,0,1,0), (1,1,1,0), (0,0,0,1), (0,1,0,1), (0,0,1,1), (0,1,1,1),\\
   			&~(1,0,0,1), (1,1,0,1), (1,0,1,1), (1,1,1,1)\},\\
   			\Delta^{c}_{N}=&\{(1,0,0,0), (1,1,0,0), (0,0,1,0), (0,1,1,0), (0,0,0,1), (0,1,0,1), (1,0,1,0), (1,1,1,0), \\ &~(1,0,0,1), (1,1,0,1), (0,0,1,1), (0,1,1,1), (1,0,1,1), (1,1,1,1)\}.
   		\end{split}
   	\end{equation*}
   	Note that $C_{D^{c}}$ is a $[4064, 4, 3556]$ linear code over $\mathbb{F}_{8}$ and it is a $4$-weight linear code. In particular, $C_{D^{c}}$ have codewords of weights $0, 3556, 3560, 3568, 3584$. Observe $C_{D^{c}}$ is a Griesmer code and hence it is a distance optimal code. Since $\frac{3556}{3584}>\frac{7}{8}$, it is a minimal code by using Lemma \ref{minimal_lemma}.
   \end{example}
   \begin{example}
   	Set $m=4$ and $L=\{1, 2\}= M=N$. Let $D=\Delta_{L}+\omega \Delta_{L}+\omega^2\Delta_{L}\subset \mathbb{F}_{8}^4$ so that $D^{c}=\big(\Delta_{L}^{c}+\omega \mathbb{F}_{2}^4+\omega^2\mathbb{F}_{2}^4\big)\bigsqcup \big(\Delta_{L}+\omega \Delta_{L}^c+\omega^2\mathbb{F}_{2}^4\big)\bigsqcup \big(\Delta_{L}+\omega \Delta_{L}+\omega^2 \Delta_{L}^c\big)$, where
   	\begin{equation*}
   		\begin{split}
   			\Delta^{c}_{L}=&\{(0,0,1,0), (1,0,1,0), (0,1,1,0), (1,1,1,0), (0,0,0,1), (1,0,0,1), (0,1,0,1), (1,1,0, 1),\\ &~(0,0,1,1), (1,0,1,1), (0,1,1,1), (1,1,1,1)\}.
   		\end{split}
   	\end{equation*}   	
   	The code $C_{D^{c}}$ is a $[4032, 4, 3528]$ octanary $2$-weight linear code. In particular, $C_{D^c}$ have codewords of weights $0, 3528, 3584$. Observe that $C_{D^c}$ is a Griesmer code and hence it is a distance optimal code. Since $\frac{3528}{3584}>\frac{7}{8}$, it is a minimal code by using Lemma \ref{minimal_lemma}.
   \end{example}
       \begin{remark}\label{theoremoverfield2ncomplimnt}
       	Let $\mathbb{F}_{2^n}=\mathbb{F}_{2}(\eta)$, where $\eta\in \mathbb{F}_{2^n}$. Consider the defining set $D=\Delta_{M_1}+\eta \Delta_{M_2}+\cdots +\eta^{n-1}\Delta_{M_n}\subset \mathbb{F}_{2^n}^m$, where $m\in \mathbb{N}$ and $\emptyset\neq M_{i}\subsetneq [m], 1\leq i\leq n$ such that $\cup_{i=1}^{n}M_{i}\subsetneq [m]$. The code $C_{D^{c}}$ is a linear code of length $2^{nm}-2^{\sum_{j=1}^n|M_{j}|}$ over $\mathbb{F}_{2^n}$.   
       \end{remark}
       Based on the computations done for codes over the fields $\mathbb{F}_{2^r}, r=1, 2, 3$ ans $4$, we conjecture the following.
       \begin{conjecture}
       	Let $\mathbb{F}_{2^n}=\mathbb{F}_{2}(\eta)$, where $\eta\in \mathbb{F}_{2^n}$. Consider the defining set $D=\Delta_{M_1}+\eta \Delta_{M_2}+\cdots +\eta^{n-1}\Delta_{M_n}\subset \mathbb{F}_{2^n}^m$, where $m\in \mathbb{N}$ and $\emptyset\neq M_{i}\subsetneq [m], 1\leq i\leq n$ such that $\cup_{i=1}^{n}M_{i}\subsetneq [m]$.
        The code $C_{D^c}$ in remark \ref{theoremoverfield2ncomplimnt} is an $m$-dimensional code with distance
       $(2^{n}-1)\times \big(2^{n(m-1)}- 2^{\sum_{j=1}^n|M_{j}|-n}\big)$ and it is a $(t+1)$-weight linear code over $\mathbb{F}_{2^n}$ provided $C_{D^{\ast}}$ is a $t$-weight linear code over $\mathbb{F}_{2^n}$. Moreover, it is a Griesmer code. In fact, $C_{D^c}$ is a minimal code if $\sum_{j=1}^n|M_{j}|\leq nm- (n+1)$.
       \end{conjecture}
     \section{Subfield code with respect to a linear code}\label{section_4}
       In this section, we study subfield codes over $\mathbb{F}_2$ with respect to codes discussed earlier.
    
        \begin{theorem}\label{subfieldthmDF8}
    	Let $L, M, N$ be nonempty subsets of $[m]$ and let $D=\Delta_{L}+\omega\Delta_{M}+\omega^2\Delta_{N} \subset \mathbb{F}_{8}^m$ so that $D^{(2)}=\{(d_1, d_3, d_2): d_1\in \Delta_{L}, d_2\in\Delta_{M}, d_3\in \Delta_{N}\}\subset (\mathbb{F}_{2}^m)^3$. Then the code $C^{(2)}_{D^{\ast}}$ is a $[2^{|L|+|M|+|N|}-1, |L|+|M|+|N|, 2^{|L|+|M|+|N|-1}]$ linear $1$-weight code over $\mathbb{F}_{2}$ and its weight distribution is recorded in Table \ref{table:2.1}. In particular, $C^{(2)}_{D^{\ast}}$ is a minimal code. Further, it is a Griesmer code and hence it is a distance optimal.
    	\begin{table}
    	  \centering
   			\begin{tabular}{|c|c|}
    				\hline
    				Hamming weight    & Number of codewords \\
    				\hline
    				$0$ & $1$\\
    				\hline
    				$2^{|L|+|M|+|N|-1}$ & $2^{|L|+|M|+|N|}-1$\\    				
    				\hline
    		\end{tabular}    	
    		\caption{Weight distribution in Theorem \ref{subfieldthmDF8}}
    		\label{table:2.1}
    	\end{table}
        If $Z_{i} =|\{(\alpha, \beta, \gamma)\in (\mathbb{F}_{2}^m)^3 : wt(c^{(2)}_{D^{\ast}}(\alpha, \beta, \gamma))=i \}|$, where $i=0, 2^{|L|+|M|+|N|-1}$, then
        \begin{equation*}       		
        	Z_{0}=2^{3m-|L|-|M|-|N|} \textnormal{ and } Z_j=2^{3m-|L|-|M|-|N|}\times  \big(2^{|L|+|M|+|N|}-1\big).
        \end{equation*}
     \begin{proof}
     	Note that the length of $C^{(2)}_{D^{\ast}}$ is $|D^{\ast}|=|{D^{(2)}}^{\ast}| = 2^{|L|+|M|+|N|}-1$. Observe that the map $c^{(2)}_{D^{\ast}}: (\mathbb{F}_{2}^m)^3\longrightarrow C^{(2)}_{D^{\ast}}$ defined by $c^{(2)}_{D^{\ast}}(x)=\big(x\cdot d\big)_{d\in D^{(2)^\ast}}$ is a surjective linear transformation. By Theorem \ref{matrixtheorem},
     	\begin{equation*}
     		C^{(2)}_{D^{\ast}}=\{c^{(2)}_{D^{\ast}}(\alpha, \beta, \gamma)=\big((\alpha, \beta, \gamma )\cdot (d_1, d_3, d_2)\big)_{d_1\in \Delta_{L}, d_2\in \Delta_{M}, d_3\in \Delta_{N}}: \alpha, \beta, \gamma \in \mathbb{F}_{2}^m\}.
     	\end{equation*}     
     Now, we have
     \begin{equation}\label{weightCd2F8}
     	\begin{split}
     		wt\big(c^{(2)}_{D^{\ast}}(\alpha, \beta, \gamma)\big)=&wt\big(c^{(2)}_{D}(\alpha, \beta, \gamma)\big)\\
     		=&|D|-\frac{1}{2}\sum\limits_{d_1\in \Delta_{L}}\sum\limits_{d_2\in \Delta_{M}}\sum\limits_{d_3\in \Delta_{N}}\big(1+(-1)^{\alpha d_1+\gamma d_2+\beta d_3}\big)\\
     		=&\frac{1}{2}|D|-\frac{1}{2}\sum\limits_{d_1\in \Delta_{L}}(-1)^{\alpha d_1}\sum\limits_{d_2\in \Delta_{M}}(-1)^{\gamma d_2}\sum\limits_{d_3\in \Delta_{N}}(-1)^{\beta d_3}\\
     		=&2^{|L|+|M|+|N|-1}\big(1-\varphi(\alpha|L)\varphi(\gamma|M)\varphi(\beta|N)\big).
     	\end{split}
     \end{equation}
     Case: (1)\\
     $wt(c^{(2)}_{D^{\ast}}(\alpha, \beta, \gamma))=0 \iff \varphi(\alpha|L)\varphi(\gamma|M)\varphi(\beta|N)=1 \iff \textnormal{ Supp}(\alpha)\cap L = \textnormal{Supp}(\gamma)\cap M= \textnormal{Supp}(\beta)\cap N=\emptyset$. Hence, $Z_0=2^{m-|L|}\times 2^{m-|M|}\times 2^{m-|N|}=2^{3m-|L|-|M|-|N|}=|\ker(c^{(2)}_{D^{\ast}})|$. By the first isomorphism theorem of groups, we have $|C^{(2)}_{D^{\ast}}|=\frac{|(\mathbb{F}_{2}^m)^3|}{|\ker(c^{(2)}_{D^\ast})|}=2^{|L|+|M|+|N|}$. Hence $\dim(C^{(2)}_{D^{\ast}})=|L|+|M|+|N|$. \\
     Case: (2)\\
     $wt(c^{(2)}_{D^{\ast}}(\alpha, \beta, \gamma))=2^{|L|+|M|+|N|-1} \iff \varphi(\alpha|L)\varphi(\gamma|M)\varphi(\beta|N)=0$.\\ Therefore, $Z_j=2^{3m-|L|-|M|-|N|}\times \big(2^{|L|+|M|+|N|}-1\big)$, where $j=2^{|L|+|M|+|N|-1}$.\\ Since $C_{D^{\ast}}^{(2)}$ is a $1$-weight code, it is a minimal code.
     Now we have
     \begin{equation*}
     	\begin{split}
     		\sum\limits_{i=0}^{|L|+|M|+|N|-1}\left\lceil \frac{2^{|L|+|M|+|N|-1}}{2^i}\right\rceil 
     		= &\big(2^{|L|+|M|+|N|}-1\big).
     	\end{split}
     \end{equation*}
     Therefore, $C^{(2)}_{D^c}$ is a Griesmer code.
     \end{proof}
    \end{theorem}
   Here are examples illustrating Theorem \ref{subfieldthmDF8}.
   \begin{example}
   	Set $m=3$ and $L=\{1, 2\}, M=\{2,3\}, N=\{2\}\subseteq [3]=\{1, 2, 3\}$. Let the defining set $D=\Delta_{L}+\omega \Delta_{M}+\omega^2\Delta_{N}\subset \mathbb{F}_{8}^3$ so that $D^{(2)}=\{(d_1, d_3, d_2): d_1\in \Delta_{L}, d_2\in \Delta_{M}, d_3\in \Delta_{N}\}$, where
   	\begin{equation*}
   		\begin{split}
   			\Delta_{L}=&\{(0,0,0), (1,0,0), (0, 1, 0), (1, 1, 0)\},\\ \Delta_{M}=&\{(0,0,0), (0,1,0), (0,0,1), (0,1,1)\},\\
   			\Delta_{N}=&\{(0,0,0), (0,1,0)\}.
   		\end{split}
   	\end{equation*}
   	Note that $C^{(2)}_{D^{\ast}}$ is a $[31, 5, 16]$ linear $1$-weight code over $\mathbb{F}_{2}$ which is distance optimal from \citep{BoundTable}. Since $C^{(2)}_{D^{\ast}}$ is a $1$-weight code, it is a minimal code.
   \end{example}
    \begin{example}
    	Set $m=2$ and $L=\{1\}, M=\{2\}, N=\{2\}\subseteq [2]=\{1,2\}$. Let the defining set $D=\Delta_{L}+\omega \Delta_{M}+\omega^2\Delta_{N}\subset \mathbb{F}_{8}^2$ so that $D^{(2)}=\{(d_1, d_3, d_2): d_1\in \Delta_{L}, d_2\in \Delta_{M}, d_3\in \Delta_{N}\}$, where
    	\begin{equation*}
    		\begin{split}
    			\Delta_{L}=&\{(0,0), (1,0)\},\\ 
    			\Delta_{M}=&\{(0,0), (0,1)\},\\
    			\Delta_{N}=&\{(0,0), (0,1)\}.
    		\end{split}
    	\end{equation*}
    	Note that $C^{(2)}_{D^{\ast}}$ is a $[7, 3, 4]$ linear $1$-weight code over $\mathbb{F}_{2}$ which is distance optimal from \citep{BoundTable}. Since $C^{(2)}_{D^{\ast}}$ is a $1$-weight code, it is a minimal code.
    \end{example}
   The following result gives parameters of subfield codes with respect to linear codes over $\mathbb{F}_{2^n}$.
    \begin{proposition}\label{subfieldTheoremF2n}
    	Let $M_i$ be nonempty subsets of $[m]$ for $1\leq i\leq n$. Suppose $D=\Delta_{M_1}+\eta \Delta_{M_2}+\cdots +\eta^{n-1}\Delta_{M_n}\subset \mathbb{F}_{2^n}^m$, where $n,m\in \mathbb{N}$. Then the code $C^{(2)}_{D^{\ast}}$ is a $1$-weight linear code over $\mathbb{F}_{2}$ of length $2^{\sum_{j=1}^{n}|M_j|}-1$ and distance $2^{\sum_{j=1}^{n}|M_j|-1}$. In particular, $C^{(2)}_{D^{\ast}}$ is a minimal code.
    	\begin{proof}
    		Observe $|D^{\ast}|=|D^{(2)^{\ast}}|=2^{\sum_{j=1}^{n}|M_j|}-1$. Therefore the length of the code $C^{(2)}_{D^{\ast}}$ with respect to the code in Proposition \ref{theoremoverfield2n} is $2^{\sum_{j=1}^{n}|M_j|}-1$.
    		Consider the map $c^{(2)}_{D^{\ast}}: (\mathbb{F}_{2}^m)^n\longrightarrow C^{(2)}_{D^{\ast}}$ defined by $c^{(2)}_{D^{\ast}}(x)=\big(x\cdot d\big)_{d\in D^{(2)^\ast}}$, which is a surjective linear transformation.\\    		
    		Now,
    		\begin{equation*}
    			\begin{split}
    				wt\big(c^{(2)}_{D^{\ast}}(x)\big)=&wt\big(c^{(2)}_{D}(x)\big)\\
    				=&|D^{(2)}|-\frac{1}{2}\sum\limits_{d\in D^{(2)}}\big(1+(-1)^{x\cdot d}\big)\\
    				=&\frac{1}{2}|D^{(2)}|-\frac{1}{2}\sum\limits_{d\in D^{(2)}}(-1)^{x\cdot d}\\ 				
    				=&2^{\sum_{j=1}^{n}|M_j|-1}\big(1-\theta\big)\textnormal{, where } \theta \textnormal{ is a product of some Boolean functions}.
    			\end{split}
    		\end{equation*}
    		Case: $\theta=0$\\
    		$wt(c^{(2)}_{D^{\ast}}(x))=0$ if and only if $\theta=1$ and an example of such $x$ for which $\theta=1$ is $x=(0,\dots, 0)$.
    		Case: $\theta=1$\\
    		$wt(c^{(2)}_{D^{\ast}}(x))=2^{\sum_{j=1}^{n}|M_j|-1}$ if and only if $\theta=0$. Since $C^{(2)}_{D^{\ast}}\neq 0$, there exists $x$ for which $\theta=0$. Thus $C_{D^{\ast}}^{(2)}$ is a $1$-weight code.
    	\end{proof}
    \end{proposition}
    Based on the computations done for subfield codes with respect to codes over $\mathbb{F}_{2^r}, r=1, 2, 3$ and $4$, we conjecture the following.
    \begin{conjecture}
    		Let $M_i$ be nonempty subsets of $[m]$ for $1\leq i\leq n$. Suppose $D=\Delta_{M_1}+\eta \Delta_{M_2}+\cdots +\eta^{n-1}\Delta_{M_n}\subset \mathbb{F}_{2^n}^m$, where $n,m\in \mathbb{N}$. Then the dimension of the code $C^{(2)}_{D^{\ast}}$ in Proposition \ref{subfieldTheoremF2n} is $\sum_{j=1}^{n}|M_j|$. Moreover, $C^{(2)}_{D^{\ast}}$ is a Griesmer code. If $Z_i=|\{x\in (\mathbb{F}_{2}^m)^n: wt(c^{(2)}_{D^{\ast}}(x))=i\}|$ for $i=0, 2^{\sum_{j=1}^{n}|M_j|-1}$, then
    	\begin{equation*}
    		Z_0=2^{nm-\sum_{j=1}^{n}|M_{j}|} \textnormal{ and } Z_j=Z_0A_j.
    	\end{equation*}
    \end{conjecture}
     Suppose $D=\Delta_{L}+\omega \Delta_{M}+\omega^2 \Delta_{N}\subset \mathbb{F}_{8}^m$ then by using Equation \ref{DcExpressionF8}, we have
     \begin{equation*}
     	D^{c}=\big(\Delta_{L}^{c}+\omega \mathbb{F}_{2}^m+\omega^2\mathbb{F}_{2}^m\big)\bigsqcup \big(\Delta_{L}+\omega \Delta_{M}^c+\omega^2\mathbb{F}_{2}^m\big)\bigsqcup \big(\Delta_{L}+\omega \Delta_{M}+\omega^2 \Delta_{N}^c\big),
     \end{equation*}
      where $\sqcup$ denotes disjoint union.\\
     If $(\alpha, \beta, \gamma)\in (\mathbb{F}_{2}^m)^3$ then we have
     \begin{equation*}
     	\begin{split}
     		wt\big(c^{(2)}_{D^c}(\alpha, \beta, \gamma)\big)=&|D^c|-\frac{1}{2}\sum\limits_{d_1\in \Delta_{L}^{c}}\sum\limits_{d_2\in \mathbb{F}_{2}^m}\sum\limits_{d_3\in \mathbb{F}_{2}^m}\big(1+(-1)^{\alpha d_1+\gamma d_2+\beta d_3}\big)\\
     		&-\frac{1}{2}\sum\limits_{d_1\in \Delta_{L}}\sum\limits_{d_2\in \Delta_{M}^{c}}\sum\limits_{d_3\in \mathbb{F}_{2}^m}\big(1+(-1)^{\alpha d_1+\gamma d_2+\beta d_3}\big)\\
     		&-\frac{1}{2}\sum\limits_{d_1\in \Delta_{L}}\sum\limits_{d_2\in \Delta_{M}}\sum\limits_{d_3\in \Delta_{N}^{c}}\big(1+(-1)^{\alpha d_1+\gamma d_2+\beta d_3}\big)\\
     		=&\frac{1}{2}\big(|D^{c}|-2^{3m}\delta_{0, \alpha}\delta_{0, \beta}\delta_{0, \gamma}\big)+\frac{1}{2}\sum\limits_{d_1\in \Delta_{L}}(-1)^{\alpha d_1}\sum\limits_{d_2\in \Delta_{M}}(-1)^{\gamma d_2}\sum\limits_{d_3\in \Delta_{N}}(-1)^{\beta d_3}\\
     		=&\frac{1}{2}\big(|D|+|D^c|-2^{3m}\delta_{0, \alpha}\delta_{0, \beta}\delta_{0, \gamma}\big)- wt\big(c^{(2)}_{D^{\ast}}(\alpha,\beta,\gamma)\big)\textnormal{, By using Equation }\ref{weightCd2F8}
     		\\=&2^{3m-1}\times \big(1-\delta_{0,\alpha}\delta_{0,\beta}\delta_{0,\gamma}\big)- wt\big(c^{(2)}_{D^{\ast}}(\alpha,\beta,\gamma)\big).
     	\end{split}
     \end{equation*}
     \begin{lemma}\label{weightrelationF23Dc}
     	Let the symbols be the same as above. For $(\alpha, \beta, \gamma)\in (\mathbb{F}_{2}^m)^3$, we have
     	\begin{equation}
     		wt\big(c^{(2)}_{D^c}(\alpha, \beta, \gamma)\big)+ wt\big(c^{(2)}_{D^{\ast}}(\alpha,\beta,\gamma)\big)=2^{3m-1}\times \big(1-\delta_{0,\alpha}\delta_{0,\beta}\delta_{0,\gamma}\big),
     	\end{equation}
     where $\delta$ denotes the Kronecker delta function.
     \end{lemma}
 
 \begin{theorem}\label{subfieldthmDcF8}
 	Suppose $L, M, N$ are nonempty subsets of $[m]$ such that at least one subset is proper. Let $D=\Delta_{L}+\omega \Delta_{M}+\omega^2\Delta_{N}\subset \mathbb{F}_{8}^m$. Then the code $C_{D^{c}}^{(2)}$ is a $[2^{3m}-2^{|L|+|M|+|N|}, 3m, 2^{3m-1}-2^{|L|+|M|+|N|-1}]$ binary $2$-weight linear code and its weight distribution is recorded in Table \ref{table:2.2}. Moreover, it is a Griesmer code and hence it is distance optimal. Further, if $|L|+|M|+|N|\leq 3m-2$ then $C_{D^{c}}^{(2)}$ is a minimal code.
 	\begin{table}
 		\begin{center}
 			\begin{tabular}{|c|c|}
 				\hline
 				Hamming weight    & Number of codewords \\
 				\hline
 				$0$ & $1$\\
 				\hline
 				$2^{3m-1}-2^{|L|+|M|+|N|-1}$ & $2^{3m-|L|-|M|-|N|}\times \big(2^{|L|+|M|+|N|}-1 \big)$\\    				
 				\hline
 				$2^{3m-1}$ & $2^{3m-|L|-|M|-|N|}-1$\\
 				\hline
 			\end{tabular}
 		\end{center}
 		\caption{Weight distribution in Theorem \ref{subfieldthmDcF8}}
 		\label{table:2.2}
 	\end{table}
 	\begin{proof} 		
 		By Theorem \ref{matrixtheorem}, $C^{(2)}_{D^c}=\{c^{(2)}_{D^c}(\alpha, \beta, \gamma)=((\alpha, \beta, \gamma)\cdot d)_{d\in ({D^{c}})^{(2)}}: \alpha, \beta, \gamma \in \mathbb{F}_{2}^m\}$. Clearly, the length of the code $C_{D^{c}}^{(2)}$ is $|D^c|= 2^{3m}-2^{|L|+|M|+|N|}$. By Lemma \ref{weightrelationF23Dc} and the weight distribution of $C^{(2)}_{D^{\ast}}$, we have the weight distribution of $C^{(2)}_{D^c}$ which is recorded in Table \ref{table:2.2}.\\
 		Now we have\\
 		\begin{equation*}
 			\begin{split}
 				\sum\limits_{i=0}^{3m-1}\left\lceil \frac{2^{3m-1}-2^{|L|+|M|+|N|-1}}{2^i}\right\rceil =&\sum\limits_{i=0}^{3m-1} \frac{2^{3m-1}}{2^i}-\sum\limits_{i=0}^{3m-1}\left\lfloor \frac{2^{|L|+|M|+|N|-1}}{2^i}\right\rfloor\\
 				= &\big(2^{3m}-1\big)-\big(2^{|L|+|M|+|N|}-1\big)\\
 				=& 2^{3m}-2^{|L|+|M|+|N|}.
 			\end{split}
 		\end{equation*}
 		Therefore, $C^{(2)}_{D^c}$ is a Griesmer code. By Lemma \ref{minimal_lemma}, we have
 		\begin{equation*}
 			\begin{split}
 				\frac{wt_{min}}{wt_{max}}=& \frac{2^{3m-1}-2^{|L|+|M|+|N|-1}}{2^{3m-1}}\\
 				=& 1-2^{|L|+|M|+|N|-3m}  		
 			\end{split}
 		\end{equation*}
 		and $1-2^{|L|+|M|+|N|-3m}>\frac{1}{2}\iff |L|+|M|+|N|\leq 3m-2$.\\ Hence, $C^{(2)}_{D^c}$ is minimal if $|L|+|M|+|N|\leq 3m-2$.
 	\end{proof}
   \end{theorem}
\begin{table}[]
	\centering
	\begin{adjustbox}{width=\textwidth}
		
		\begin{tabular}{|c|c|c|c|c|c|c|}
			\hline
			Reference &	$q$-ary & Result  & Defining set & $[n,k,d]$-code & \#Weight & Bound  \\
			\hline
			\multirow{8}{*}{\citep{Wu_Li}}&\multirow{4}{*}{4}&Proposition $4.2$&\multirow{2}{*}{$\Delta_{A}+\omega\Delta_{B}$}&$[2^{|A|+|B|}-1, |A\cup B|, 2^{|A|+|B|-1}]$&2& \\
			\cline{3-3}\cline{5-7}
			& &Proposition $4.7$& &$[(2^{|A|}+2^{|B|}-2^{|A\cap B|})^2-1, |A\cup B|, d]$&$\leq 10$ & \\
			\cline{3-3}\cline{4-7}
			& & Theorem $4.4$&\multirow{2}{*}{$\mathbb{F}_{4}^m\setminus (\Delta_{A}+\omega\Delta_{B})$}&$[4^m-2^{|A|+|B|}, m, 3\times 2^{2m-2}-3\times 2^{|A|+|B|-2}]$&3& Griesmer\\
			\cline{3-3}\cline{5-7}                                            
			& & Theorem $4.10$& &$[4^m-(2^{|A|}+2^{|B|}-2^{|A\cap B|})^2,m]$&$\leq 11$& \\
			\cline{2-7}
			
			&\multirow{5}{*}{2}&Proposition $5.1$&\multirow{2}{*}{$\Delta_{A}+\omega\Delta_{B}$}&$[2^{|A|+|B|}-1, |A|+ |B|, 2^{|A|+|B|-1}]$&1& \\
			\cline{3-3}\cline{5-7}
			& & Theorem $5.2$& &$[2^{2m}-2^{|A|+|B|}, 2m, 2^{2m-1}-2^{|A|+|B|-1}]$&$2$ &Griesmer \\
			\cline{3-3}\cline{4-7}
			& & Proposition $5.3$&\multirow{3}{*}{$\Delta+\omega\Delta$}&$[(2^{|A|}+2^{|B|}-2^{|A\cap B|})^2-1, 2|A\cup B|]$&$\leq 10$& \\
			\cline{3-3}\cline{5-7}                                            
			& & Theorem $5.5$& &$[(2^{|A|}+2^{|B|}-2^{|A\cap B|})^2-1, (2^{|A|}+2^{|B|}-2^{|A\cap B|})^2-1-2|A\cup B|, 3]$& &Sphere packing \\
			\cline{3-3}\cline{5-7}
			& & Proposition $5.6$& &$[4^m-(2^{|A|}+2^{|B|}-2^{|A\cap B|})^2, 2m]$& $\leq 11$&    \\              
			\hline
			\multirow{3}{*}{\citep{Zhu_Wei}} & \multirow{3}{*}{4} & Theorem $3.1$ & $\mathbb{F}_{4}^m\setminus (\Delta_{A}+\omega \Delta_{B})$ & $[(2^m-2^{|A|})2^{|B|}, m, 3(2^{m+|B|-2}-2^{|A|+|B|-2})]$ & 5 & \\
			\cline{3-7}
			&                    & Corollary $3.2$ & $\mathbb{F}_{2}^m+\omega \Delta_{B}$ & $[2^{m+|B|}, m, 2^{m+|B|-1}]$ & 2 & \\
			\cline{3-7}
			&                    & Corollary $3.3$& $\mathbb{F}_{2}^m\setminus\Delta_{A}+\omega\mathbb{F}_{2}^m$ & $[(2^m-2^{|A|})2^m, m, 3(2^{2m-2}-2^{|A|+m-2})]$  & 2 & Griesmer\\
			
			\hline
			\multirow{8}{*}{\citep{Hyun_Lee}} &\multirow{8}{*}{2} & \multirow{2}{*}{Lemma $7$} & $\Delta^{\ast}$ & $[2^{|A|}-1, |A|, 2^{|A|-1}]$ & 1 & Griesmer \\
			\cline{4-7}
			&                   & & \multirow{3}{*}{$\mathbb{F}_{2}^m\setminus \Delta$} & $[2^{m}-2^{|A|}, m, 2^{m-1}-2^{|A|-1}]$ & 2 & Griesmer \\
			\cline{3-3}  \cline{5-7}
			&                   &    Ex. $10$               &  & $[2^{m-1}, m, 4]$ &  & Sphere Packing \\
			\cline{3-3}\cline{5-7}
			&                   &       Corollary $21$            &  & $[2^m-2\sum_{i=1}^{s}2^{|A_i|-1}+s-1, m, 2^{m-1}-\sum_{i=1}^{s}2^{|A_i|-1}]$ &  & Griesmer\\
			\cline{3-7}
			&                   & \multirow{2}{*}{Lemma $26$} &\multirow{2}{*}{$\Delta^{\ast}$} & $[2^{|A_1|}+2^{|A_2|}-2, |A_1\cup A_2|, 2^{|A_1|-1}]$ & 3 &  \\
			\cline{5-7}
			&                   & &  & $[2^{|A_1|}+2^{|A_2|}-2^{|A_1\cap A_2|}-1, |A_1\cup A_2|, 2^{|A_1|-1}]$ & 4 & \\
			\cline{3-7}
			&                  & \multirow{2}{*}{Theorem $27$} &\multirow{2}{*}{$\mathbb{F}_{2}^m\setminus \Delta$} & $[2^m-2^{|A_1|}-2^{|A_2|}+1, m, 2^{m-1}-2^{|A_1|-1}2^{|A_2|-1}]$ & 3 or 4 & Griesmer \\
			\cline{5-7}
			&                   & &  & $[2^m-2^{|A_1|}-2^{|A_2|}+2^{|A_1\cap A_2|}-1, m, 2^{m-1}-2^{|A_1|-1}-2^{|A_2|-1}]$ & 4 or 5 & \\
			\hline
			\multirow{2}{*}{\citep{Wu_Lee}} & \multirow{2}{*}{2} &Theorem $5$ & \multirow{2}{*}{$\Delta_{A}\setminus \Delta_{B}$} & $[2^{|A|}-2^{|B|}, |A|, 2^{|A|-1}-2^{|B|-1}]$ & 2 & Griesmer \\
			\cline{3-3} \cline{5-7}
			& &Theorem $6$ &  & $[2^{|A|}-2^{|B|}, 2^{|A|}-2^{|B|}-|A|, 3 \text{ or }4]$ &  &  \\
			\hline
			\multirow{5}{*}{\citep{Hyun_Kim_Na}} & \multirow{5}{*}{$p$} & Theorem $4.1$& \multirow{5}{*}{$\mathbb{F}_{p}^m\setminus \Delta$} & $[p^m-r-1, m, (p-1)p^{m-1}-r]$ & 2 & Griesmer \\
			\cline{3-3} \cline{5-7}
			&  & Theorem $4.4$ &  & $[p^m-2r-2, m, (p-1)p^{m-1}-2r-1]$ & 4 & Griesmer\\
			\cline{3-3}	\cline{5-7}
			&  & Theorem $4.7$ &  & $[p^m-3(r+1), m, (p-1)p^{m-1}-3r-2]$ & 5 & Griesmer\\
			\cline{3-3} \cline{5-7}
			&  & Theorem $4.11$ &  & $[p^m-(r+1)(p-1), m, (p-1)p^{m-1}-(r+1)p+2r+1]$ & 4 & Griesmer\\
			\cline{3-3} \cline{5-7}
			&  & Theorem $4.14$ &  & $[p^m-(r+1)(p-2), m, (p-1)p^{m-1}-(r+1)p+3r+1]$ & 5 & Griesmer\\
			\hline
			
		\end{tabular}
		
	\end{adjustbox}
	\caption{Linear codes from Simplicial complexes before this article}
	\label{table:2.3} 		
\end{table}
\begin{table}[]
	\centering
	\begin{adjustbox}{width=\textwidth}
		
		\begin{tabular}{|c|c|c|c|c|c|c|}
			\hline
			$q$-ary  & 	Result &	Defining set & $[n,k,d]$-code & \#Weight & Distance optimal & Minimal  \\
			\hline
			\multirow{4}{*}{8}	& Proposition \ref{DtheoremF8}  & $(\Delta_{L}+\omega\Delta_{M}+\omega^2\Delta_{N})\setminus \{0\}$	& $[2^{|L|+|M|+|N|}-1, |L\cup M\cup N|, 2^{|L|+|M|+|N|-1}]$ & 3 &  &  \\
			\cline{2-7}
			&	Theorem \ref{CorCDF8}  &$(\Delta_{L}+\omega\Delta_{L}+\omega^2\Delta_{L})\setminus \{0\}$	& $[2^{3|L|}-1, |L|, 7\times 2^{3(|L|-1)}]$ & 1 & Yes & Yes \\
			\cline{2-7}
			&	Theorem \ref{theoremF8Dc}  & $\mathbb{F}_{8}^m\setminus(\Delta_{L}+\omega\Delta_{M}+\omega^2\Delta_{N})$ & $[2^{3m}-2^{|L|+|M|+|N|}, m, 7\times 2^{3(m-1)}- 7\times 2^{|L|+|M|+|N|-3}]$ & 4 & Yes & Yes, if $|L|+|M|+|N|\leq 3m-4$ \\ 
			\cline{2-7}
			&	Theorem \ref{CorCDcF8}  & $\mathbb{F}_{8}^m\setminus(\Delta_{L}+\omega\Delta_{L}+\omega^2\Delta_{L})$ 	& $[2^{3m}-2^{3|L|}, m, 7\times 2^{3(m-1)}-7\times 2^{3(|L|-1)}]$ & 2 & Yes & Yes, if $3(m-|L|)-4\geq 0$ \\
			\hline
			\multirow{2}{*}{2} &	Theorem \ref{subfieldthmDF8}  & $(\Delta_{L}+\omega\Delta_{M}+\omega^2\Delta_{N})\setminus\{0\}$ 	& $[2^{|L|+|M|+|N|}-1, |L|+|M|+|N|, 2^{|L|+|M|+|N|-1}]$ & 1 & Yes & Yes \\ \cline{2-7}
			&	Theorem \ref{subfieldthmDcF8} & ${\mathbb{F}_{8}^m\setminus (\Delta_{L}+\omega\Delta_{M}+\omega^2\Delta_{N})}$ 	& $[2^{3m}-2^{|L|+|M|+|N|}, 3m, 2^{3m-1}- 2^{|L|+|M|+|N|-1}]$ & 2 & Yes & Yes, if $|L|+|M|+|N|\leq 3m-2$ \\
			\hline  						
		\end{tabular}
		
	\end{adjustbox}
	\caption{Linear codes from Simplicial complexes in this article}
	\label{table:2.4}
	
\end{table}

   Here are examples illustrating Theorem \ref{subfieldthmDcF8}.
   \begin{example}
   	Set $m=3$ and $L=\{1, 2\}, M=\{2,3\}, N=\{2\}\subseteq [3]=\{1, 2, 3\}$. Let the defining set $D=\Delta_{L}+\omega \Delta_{M}+\omega^2\Delta_{N}\subset \mathbb{F}_{8}^3$ so that $D^{c}=\big(\Delta_{L}^{c}+\omega \mathbb{F}_{2}^3+\omega^2\mathbb{F}_{2}^3\big)\bigsqcup \big(\Delta_{L}+\omega \Delta_{M}^c+\omega^2\mathbb{F}_{2}^3\big)\bigsqcup \big(\Delta_{L}+\omega \Delta_{M}+\omega^2 \Delta_{N}^c\big)$, where
   	\begin{equation*}
   		\begin{split}
   			\Delta_{L}^c=&\{(0,0,1), (1,0,1), (0, 1, 1), (1, 1, 1)\},\\ \Delta_{M}^c=&\{(1,0,0), (1,1,0), (1,0,1), (1,1,1)\},\\
   			\Delta_{N}^c=&\{(1,0,0), (1,1,0),(0,0,1), (0,1,1), (1,0,1), (1,1,1)\}.
   		\end{split}
   	\end{equation*}
   	Note that $C^{(2)}_{D^{c}}$ is a $[480, 9, 240]$ linear code over $\mathbb{F}_{2}$ and it is a $2$-weight linear code. In particular, $C^{(2)}_{D^{c}}$ have codewords of weights $0, 240, 256$. Observe $C^{(2)}_{D^{c}}$ is a Griesmer code and hence it is a distance optimal code. Since $\frac{240}{256}>\frac{1}{2}$, it is a minimal code by using Lemma \ref{minimal_lemma}.
   \end{example}
   \begin{example}
   	Set $m=2$ and $L=\{1\}, M=\{2\}, N=\{2\}\subseteq [2]$. Let the defining set $D=\Delta_{L}+\omega \Delta_{M}+\omega^2\Delta_{N}\subset \mathbb{F}_{8}^2$ so that $D^{c}=\big(\Delta_{L}^{c}+\omega \mathbb{F}_{2}^2+\omega^2\mathbb{F}_{2}^2\big)\bigsqcup \big(\Delta_{L}+\omega \Delta_{M}^c+\omega^2\mathbb{F}_{2}^2\big)\bigsqcup \big(\Delta_{L}+\omega \Delta_{M}+\omega^2 \Delta_{N}^c\big)$, where
   	\begin{equation*}
   		\begin{split}
   			\Delta_{L}^c=&\{(0,1), (1,1)\},\\ \Delta_{M}^c=&\{(1,0), (1,1)\},\\
   			\Delta_{N}^c=&\{(1,0), (1,1)\}.
   		\end{split}
   	\end{equation*}
   	Note that $C^{(2)}_{D^{c}}$ is a $[56, 6, 28]$ linear code over $\mathbb{F}_{2}$ and it is a $2$-weight linear code. In particular, $C^{(2)}_{D^{c}}$ have codewords of weights $0, 28, 32$. The code $C^{(2)}_{D^c}$ is optimal from \citep{BoundTable}. Since $\frac{28}{32}>\frac{1}{2}$, it is a minimal code by using Lemma \ref{minimal_lemma}.
   \end{example}

     Now we conjecture that the relation in Lemma \ref{weightrelationF23Dc} get extended for subfield codes with respect to codes over any finite field of characteristic $2$.
     \begin{conjecture}
     	With the notations as above. For $x=(x_1,\dots, x_n)\in (\mathbb{F}_{2}^m)^n$, we have
     	\begin{equation}\label{weightrelationF2nDc}
     		\begin{split}
     			wt\big(c^{(2)}_{D^c}(x)\big)+ wt\big(c^{(2)}_{D^{\ast}}(x)\big)=2^{nm-1}\times \big(1-\delta_{0,x_1}\delta_{0, x_2}\cdots\delta_{0, x_n}\big).
     		\end{split}
     	\end{equation}
     \end{conjecture}
     \begin{remark}\label{subfieldTheoremF2nComplement}
     	Let $\mathbb{F}_{2^n}=\mathbb{F}_{2}(\eta)$, where $\eta\in \mathbb{F}_{2^n}$. Consider the defining set $D=\Delta_{M_1}+\eta \Delta_{M_2}+\cdots +\eta^{n-1}\Delta_{M_n}\subset \mathbb{F}_{2^n}^m$, where $m\in \mathbb{N}$ and $\emptyset\neq M_{i}\subset [m], 1\leq i\leq n$. Let $M_i\subsetneq$ for some $i, 1\leq i\leq n$. The code $C^{(2)}_{D^{c}}$ is a linear code of length $2^{nm}-2^{\sum_{j=1}^{n}|M_j|}$ over $\mathbb{F}_{2}$.
     \end{remark}
      Based on the computations done for subfield codes with respect to codes over $\mathbb{F}_{2^r}, r=1, 2, 3$ and $4$, we conjecture the following.
     \begin{conjecture}
     	With the notations as above, the code $C^{(2)}_{D^c}$ in remark \ref{subfieldTheoremF2nComplement} is an $nm$-dimensional code with distance
     	$2^{nm-1}-2^{\sum_{j=1}^{n}|M_j|-1}$ over $\mathbb{F}_{2}$ and it is $2$-weight linear code. Moreover, $C^{(2)}_{D^c}$ is a Griesmer code. Further, it is a minimal code if $\sum_{j=1}^{n}|M_j|\leq nm-2$.
    \end{conjecture}
            	
      \section{Conclusion}\label{section_5}
      In this manuscript, we used simplicial complexes to construct linear codes over $\mathbb{F}_{2^3}$. We study the algebraic structure of these codes and the corresponding subfield codes over $\mathbb{F}_{2}$. We produce five infinite families of linear codes that are distance optimal. The weight distributions are obtained for the binary codes considered in this manuscript, and boolean functions are used in these computations. Moreover, we obtain sufficient conditions for some of these codes to be minimal. We give a few examples to illustrate our results. Further, we partially extend these results to codes over any finite field of characteristic $2$, and propose a few conjectures. For the reader's convenience, we include lists of recent works on linear codes with the help of simplicial complexes and our work, respectively, in Table \ref{table:2.3} and \ref{table:2.4}.\par
      In future, apart from proving the conjectures, one can find weight distributions of octanary codes considered in this manuscript by devising an elegant method.

	
\end{document}